# Intrusion Detection using Network Traffic Profilingand Machine Learning for IoT

Joseph Rose\*, Matthew Swann\*, Gueltoum Bendiab\*, Stavros Shiaeles\*, Nicholas Kolokotronis†

\*Cyber Security Research Group, University of Portsmouth, PO1 2UP,
Portsmouth, UKjoseph.rose@port.ac.uk, matthew.swann@port.ac.uk,
gueltoum.bendiab@port.ac.uk, sshiaeles@ieee.org
‡Department of Informatics and Telecommunications, University of Peloponnese
22131 Tripolis, Greece. nkolok@uop.gr

*Abstract*—The rapid increase in the use of IoT devices brings many benefits to the digital society, ranging from improved efficiency to higher productivity. However, the limited resources and the open nature of these devices make them vulnerable to various cyber threats. A single compromised device can havean impact on the whole network and lead to major security and physical damages. This paper explores the potential of using network profiling and machine learning to secure IoT against cyber-attacks. The proposed anomaly-based intrusion detection solution dynamically and actively profiles and monitors all net- worked devices for the detection of IoT device tampering attemptsas well as suspicious network transactions. Any deviation from the defined profile is considered to be an attack and is subjectto further analysis. Raw traffic is also passed on to the machine learning classifier for examination and identification of potential attacks. Performance assessment of the proposed methodology is conducted on the Cyber-Trust testbed using normal and malicious network traffic. The experimental results show thatthe proposed anomaly detection system delivers promising resultswith an overall accuracy of 98.35% and 0.98% of false-positive alarms.

*Index Terms*—Machine Learning, Intrusion Detection System, Security, Internet of Things, network profiling.

## I. Introduction

IoT devices are virtually everywhere today and have been increasingly deployed to critical infrastructure sectors suchas power grids, healthcare, and industrial control systems.This integration created new entry points for the network and therefore introduced an increasing security risk [1], [2].A single compromised device can provide a foothold to internal networks and expose companies and infrastructureto major security breaches, including theft of valuable and sensitive information such as financial records and access credentials. More destructive malware, such as ransomware, can even cause the failure of medical and military equipment, endangering life or allowing breaches of national security,which is a big challenge for the whole public sector and research community [3]. These security threats should be identified before causing any type of loss or damage to the organisations. In this context, Intrusion Detection Systems(IDS) are commonly used by organisations to monitor their networks and identify possible malicious activity. However, attackers are continuously developing new attack strategies to obfuscate their attacks and avoid detection. This has been proved by the increasing number of IoT-based attacks reported annually [4], [5].

Thus, developing effective, efficient, and adaptive IDS is an active research area that researchers have been working on for decades. A large number of studies have attempted to design new IDS tailored specifically for the IoT ecosystem needs. However, most of them still need improvements in terms of scalability, detection accuracy, false alarms rate and energy consumption [6]. Also, most of the existing solutions are ineffective in detecting unknown and new versions of attacks for which do not exist signatures or predefined patterns, also known as zero-day exploits [2], [6]. To overcome these issues, this paper presents a novel intrusion detection framework based on network traffic profiling and machine learning. The proposed system involves two main components for the detection of potential known and unknown attacks on the networks level as well as on a per device basis. The first component is the profiling service which is used to perform network profiling and vulnerability assessment/identification of systems that are situated on the local network. This component has two main functionalities. First, it automatically scans all the connected devices on the locally available network for potential common vulnerabilities and currently running services. The second function is the calculation of out of bound network profile behaviours by the continual monitoring of the network traffic flow from each device across the network. Every deviation from the defined profile is considered a potential attack and is subject to further analysis. The second component is the IDS system, which is an integration of binary visualisation techniques used for malware detection alongside Suricata's network signature- based detection. Alerts generated by Suricata are sent

along- side predictive alerts of malicious network patterns which are predicted by our ML module. Our method is based on the Malware-Squid intrusion detection technique proposed in [1]. The main contribution of this paper is the integration of this technique into the Suricata IDS.

The performance evaluation of the proposed technique is carried out using our training and testing datasets. The testing data is created using the Cyber-Trust [7] testing network, whereas, the training dataset is created from both normal and malicious traffic, from pre-existing public datasets. The datasets are publicly available on the IEEE DataPort website [8]. These datasets contain both the packet captures of the network attack scenarios carried out to test IDS and profiling system as well as the images used to train the machine learning module itself. The experimental results show that the proposed IDS framework can achieve high accuracy (98.35%) with low false alarm rates (0.98%). The conducted experiments also proved the capability of our system to detect malicious traffic generated from zero-day attacks.

The rest of the paper is organised as follows: Section 2 discusses recent works in the domain. Section 3 presents the proposed IDS methodology. This section also describes the created network traffic dataset on the Cyber-Trust[1] testbed. Experimental design and results are discussed in Section 4. Finally, Section 5 concludes the paper along with some future directions.

## II. Related work

In recent years, several studies have attempted to design newIDS tailored specifically for the IoT ecosystem. In [6], authorsproposed a literature review of existing intrusion detection solutions in IoT environments. In this study, the authors proposeda classification of the existing solutions based on different features, including the intrusion detection techniques, locationof the IDS in the network, evaluation techniques, and typeof attacks. Also, the paper reviewed and deeply analysed the open issues of numerous highly developed intrusion detection systems in IoT (over 40 studies). The study concluded that existing intrusion detection solutions for IoT networks still needs improvements in terms of scalability, detection accuracy,true positive rate and energy consumption. In face of the limitations of signature-based techniques, Machine Learning (ML) has recently received considerable attention for its abilityto accurately detect intrusions and therefore reduce the false positive by detecting unknown or modified attacks.

In the same context, authors in [2] investigated the effective-ness of different machine learning algorithms in securing IoT devices against DoS attacks. This study tried to suggest appropriate methods for developing IDSs using ensemble learning for IoT applications. The assessed classifiers are Random Forest (RF), AdaBoost (AB), Extreme gradient boosting (XGB), Gradient boosted machine (GBM), and extremely randomizedtrees (ETC). In [9], authors proposed an Intrusion Detection System (IDS) framework to detect Dos/DDoS attacks in IoT ecosystem. The proposed IDS includes a monitoring system and the Suricata IDS for pattern matching and intrusions detection. The IDS probes sniff the packets and send themto the IDS, through the powerful packet manipulation pro- gram" Scapy" and a virtual interface, for further analysis.Penetration testing tool 'Scapy' is used to assess the performance of the proposed IDS. However, the authors did not provide information about signature database management andupdates. Also, there are no simulation results in support ofIDS performance. In another work [3], authors proposed a technique based on the Online Sequential Extreme Learning Machine (OS-ELM) for intrusion detection. This work used the alpha and beta profiling methods o reduce the time complexity when irrelevant features are discarded from the trainingdataset. The performance evaluation of the proposed solution is performed using the standard NSL-KDD 2009 (Network Security Laboratory-Knowledge Discovery and Data Mining) dataset. This solution achieved an overall accuracy rate of 98.66% with a false positive rate of 1.74% and a detectiontime of 2.43 seconds.

In [10], the authors proposed a new IDS mechanism todetect intrusions in IoT networks, focusing on DDoS attacks. The proposed technique maintains traditional Greylist and blacklist to control access to the network. The Greylist is updated every 40 s while the blacklist is updated every 300 s. The performance evaluation of this IDS is done on the ContikiOS and COOJA Simulator. The simulation results show highervalues of true positives, false positives, and recovery time,which makes it non-suitable for large IoT networks wherethe number of users is important. Similarly, work in [11]focused on the application of specific framework models to integrate machine learning components for flow-based traffic monitoring. This work placed the machine learning SDN controller at the centre of each network control transaction and used decision-making logic to spot inconsistencies or abnormalities to the traffic flows that pass through the SDN. This anomaly-based model is placed on top of the SDN and works in real-time to detect abnormalities. Notably missing from this model is further identification or informed action, just the detection of anomalies. This feeds into the detection rates, depending on the model used, of up to 95.16% with a false positive rate over 2.49%.

---

[1] https://cyber-trust.eu/

Work in [12] explored contemporary research and literature to compare the speed, accuracy and false-positive rates of machine learning components integrated into traditional and fully machine-learning-based IDSs. The main objective of this comparison is to check the effectiveness of different training sets for anomaly detection. The authors noted that the speed and accuracy of most current solutions are dependants on the pre-training data, as is the case with most supervised machine learning systems. However, the study noted that the data set that has been used for the majority of these systems, the KDD Cup 99' data set [13], is imbalanced and introduces limiting factors into IDS systems based on it. This is since it does not determine how accurate background traffic is to the malicious traffic for the data set. Moreover, KDD Cup 99' dataset is considered old and obsolete; thus, another purpose of this paper expects of proposing a new system is to provide the community with some recent datasets of malicious traffic which can be downloaded from [8].

### III. Proposed Methodology

Fig. 1. presents a high-level architecture of the proposed IDS framework which is composed of two main components, the network profiling service, and the IDS system. As shown in the figure, the network profiling component is located in the IoT gateway for data collection and communication. Given the additional computational requirements, this component may be relocated on a separate hardware device but closely connected to the IoT gateway (bridged with the gateway). This allows the component to access and profile the LAN network that the gateway resides on, as well as the inter-device connections. The getaway is also running a light version of the SuricataIDS (without machine learning) and smart home specific signatures (modified version of snort rules) to monitor packets on the local network and detect known malicious threats. Whereas the ML IDS component is deployed at the ISP Cloud as it requires more resources due to the ML component. The main objective of this component is the detection of potential known and unknown cyber-attacks on the networks level as well as at the device levels.

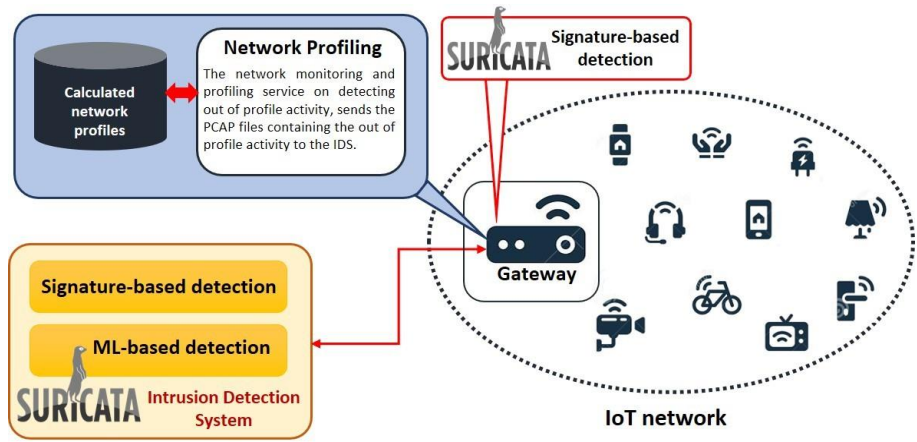

Fig. 1. High level architecture of the proposed system.

#### A. Network profiling component

The behaviour of the network profiling component is divided into two distinct parts, firstly is the functionality to automatically scan connected devices on the locally available network for potential common vulnerabilities and currently running services. The vulnerability information is primarily sourced from a routinely updated listing from the publicly available CVE Mitre database [14]. These common vulnerabilities are mapped to the available network services which in turn are discovered through network port scanning techniques such as those provided by Nmap. Once the list of potential vulnerabilities of each device is collated, each device is profiled utilising information relating to each device. This includes routing information, the reported hostname, network flow and topology; this information is then provided to external components to digest this information through a REST API framework. The second main characteristic of the networkprofiling services is the ability to calculate out of bound network profile behaviour, this is calculated by the continual monitoring of the network traffic flow from each device acrossthe network. It utilises rate informed heuristic profiling tocreate an expected throughput pattern for each device on the LAN that it is connected to. This profile is then compared against three different predefined profiles:

1) Hourly Profile (H) - A profile of the network that isinformed by a packet capture that is refreshed hourly.
2) Daily Profile (D) - A profile of the network that isinformed by a packet capture that is refreshed daily
3) Weekly Profile (W) - A profile of the network that isinformed by a packet capture that is refreshed weekly

The objective of utilising different profiles separated and refreshed by period is to provide a more accurate map of the network conditions that a device would experience over time, this increases profile accuracy and makes the system more adaptable to variable network conditions and varied device usage. The rate metric for these captures is calculated as follow:

$$RM = \frac{n}{t} \qquad (1)$$

Where $n$ is the total number of bytes transmitted and $t$ is the time of the capture. The component is then able to take periodic network captures of the LAN traffic from the gateway, this newcapture is then run through the same profiling system as the timed profile captures, and a new rate metric is calculated. A percentage difference is calculated, comparing the rate profile of the new capture to each timed profile:

$$\Delta = \frac{CRM}{tPRM} \times 100 \qquad (2)$$

Where Delta ($\Delta$) is the percentage difference between $CRM$, the calculated rate metric and $PRM$, the profile rate metric. If the delta value passes over a threshold value that can be configured per implementation depending on network volatility, the network activity of the device is flagged as out of profile and a re-scan of the network is initiated to re-scan for any possible actively exploited attack surface on the network. This process is fast, but minimal in terms of network impact and will not degrade network performance, even on a small network as the scale of the scan will increase or decrease in intensity automatically depending on scan timings and throughput. This threshold can be raised or lowered depending on the network's volatility.

If scanning is to frequent, the threshold can be increased on a busy, variable load networkfor example. The traffic capture, stored in PCAP format, that the network profiling component uses to calculate and profile each device can then be transferred to the machine learning component to check the traffic for patterns that could indicate malicious traffic, including active attacks or ongoing exploitation. This can then be used to inform mitigation actionsacross the affected network.

*B. IDS system*

The intrusion detection system uses both signature-based and anomaly-based detection techniques for the detection of potential known and unknown cyber security threats (Fig. 1). The anomaly-based detection technique is based on the Malware-Squid approach proposed in [1]. In this approach,the machine learning module makes use of the Hilbert space- filling curve as its main clustering algorithm, this is achieved by assigning specific colours to each byte as it's convertedinto a 2D image. This conversion is performed on each byte depending on its ASCII character reference:

- **blue** for printable characters
- **green** for control characters
- **red** for extended characters
- **black** for the null character, or 0x00
- **white** for the non-breaking space, or 0xFF

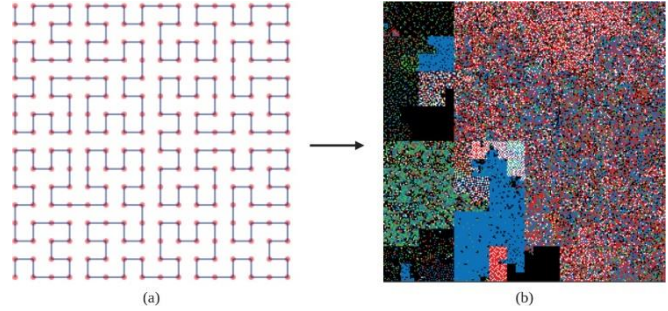

Fig. 2. (a) The Hilbert space-filling curve mapping and (b) the output 2D image.

These generated byte arrays are then processed using the Hilbert algorithm, transforming them into images that retain optimal locality for pattern recognition, allowing them to be processed by machine learning image classification models [5],where we used samples of malicious network traffic captures totrain our machine learning module. For performance reasons, multiple packets chunks are created and forward to the visual representation tool to convert them into a 2D image.

*C. Data collection and processing*

Before the tests are performed, the ML module was initiallyconditioned by a data collection of normal and malicious PCAP data. However, the overall process of the machine learning algorithm training is carried out incrementally, each time new malware traffic is found, that is, without ignoringthe information identified in earlier training phases. This wouldgreatly improve the detection accuracy of the machine learning module. The initial dataset created for the training of the machine learning module consists of more than 800 samples that were sourced from multiple malware traffic analysis repositories [15]– [17]. The dataset is publicly available on the open-access IEEE DataPort website [8]. Fig. 3 shows the percentage of malicious traffic samples in the training dataset.

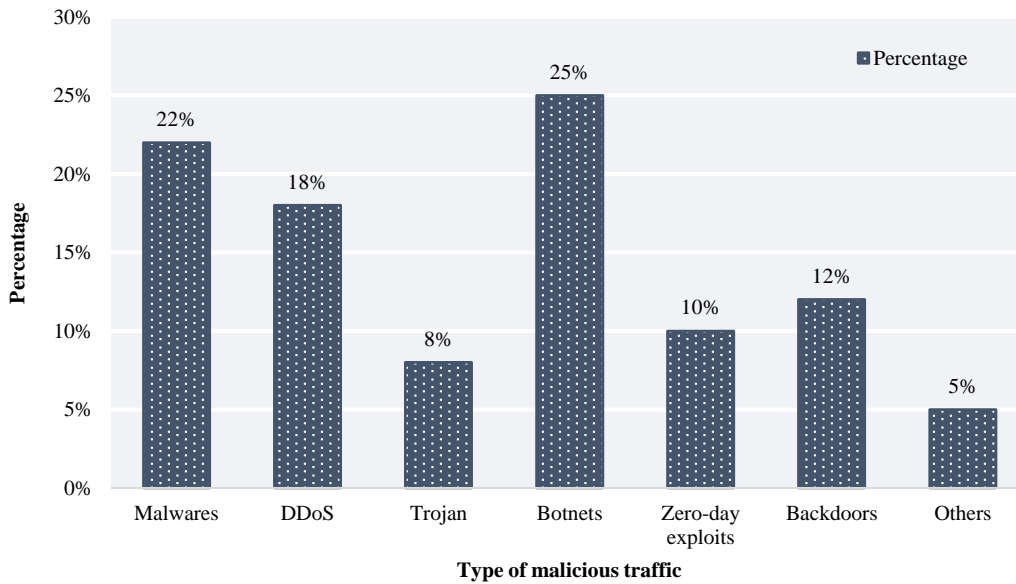

Fig. 3. Malicious traffic sample percentage according to type of malware.

To test our framework, we have created our collection of PCAP files provided by real malware traffic in the Cyber-Trusttestbed. More precisely, malicious PCAP files were created from different real-world attack scenarios, including the MiraiBotnet [18], BlackEnergy Botnet [19], Zeus Botnet [20], as well as attack replay scenario which consisted of several attack types such as a Java-RMI Backdoor [21], distcc exec backdoor [22], Web Tomcat Exploit [23] and Hydra Bruteforceattacks [24] to name a few. Full list is available in Table I.The most important attack scenario consisted of a simulated zero-day attack whereby network signatures relating to the VSFTDP [25] attack scenario was removed from the SuricataIDS and thus measured by alerts solely generated from the machine learning module. The PCAP files were generatedby running live demos of each attack scenario and recording inter-device network communication using tcpdump [26]. While normal network traffic file captures have been obtainedfrom uninfected device traffic by conducting regular network- intensive activities, such as copying both text and media files across the network, as well as running SSH sessions, streamingmedia content, and API contact, all of which may be assumed to be contained in a normal smart home network. Once these PCAP files have been stored, they can be replayed by the respective devices using TCPreplay [27] to re-transmit thecaptured packets across the network in real-time. This helpsus to track and document the responses of the IDS systemsto the actual attack data and to measure the efficacy of the detection operation for each attack scenario.

## IV. System Implementation and Testing

### A. Experimental Setup

The experiments were performed in a virtualised smart home environment in the Cyber-Trust testbed [7]. The smart home (SOHO) environment has been implemented by using several virtualised devices, divided into small groups where a separate Ubuntu VM is acting as the gateway for the SM, as depicted below (Figure 4). This VM incorporates the network profiling service and a light version of the Suricata IDS (i.e., without the ML module). Conceptually, this component may reside on the smart home gateway for data collectionand communication or given the additional computational requirements, it may be relocated on a separate hardware device but closely connected to the smart gateway. This allowsthe component to access and profile the LAN network thatthe gateway resides on, as well as the inter-device traffic as referenced in section three. The network profiling component is dockerized to allow for ease of deployment and logging.This comes with the added benefits in the form of container resource optimisation and improved execution speed, on a loaddependant basis.

The IDS system is deployed as a single VM at the ISP level, running the Suricata IDS and the anomaly detectionmachine learning module based on Debian GNU/Linux 10.2. For the SOHO virtualised devices, the following OSes were used in VM or dockerized form, Ubuntu 14/16/18.04, Windows XP-SP3, Windows 7 and Android. The smart home network configuration is done via the gateway VM whichis assigned two Interface Cards, from here we control the network assignments for both WAN and LAN traffic. The interface card eth0 is referenced as NIC1 (172.16.4.1/24) and has Internet connectivity (WAN). While the second interface eth1 is referenced as NIC2 (192.168.1.1/26) and acts as agateway IP for the smart home isolated network (LAN).

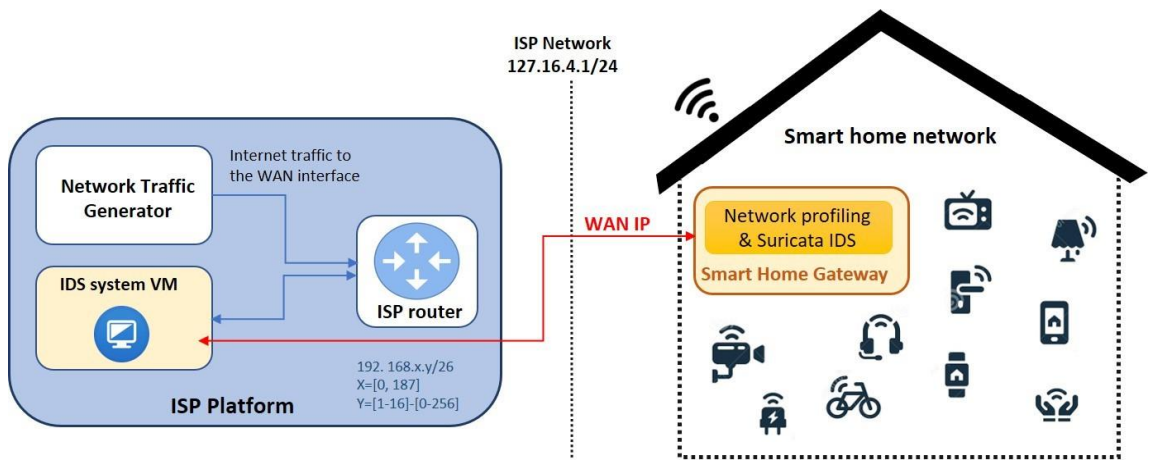

Fig. 4. Implemented Testbed

## B. Performance evaluation metrics

The performance evaluation parameters used to investigate the results of our system are accuracy (AC), false alarm or False Positive Rate (FPR), Precision (P), Recall (R) and F1- score (F1). The accuracy measures the proportion of the total number of correct classifications as in the following formula (Equation 3):

$$A = \frac{TP+TN}{TP+FP+TN+FN} \quad (3)$$

Where normal traffic represents positive instances while malicious represents negative instances. True Positive (TP) is the number of instances that have been correctly classified as legitimate. False Positive (FP) is the number of malicious instances that have been incorrectly classified as normal. True Negative (TN) is the number of samples of malicious traffic that have been correctly classified as anomalous. False Negative (FN) is the number of normal PCAP files that have been incorrectly classified as anomalous instances.

$$False\ Positive\ (FP) rate = \frac{FP}{FP+TN} \quad (4)$$

Precision (P) provides the percentage of positively classified samples that are truly positive.

$$P = \frac{TP}{TP+FP} \quad (5)$$

The recall represents the number of normal samples that were correctly classified.

$$R = \frac{TP}{TP+FN} \quad (6)$$

F-score is a weighted average between precision and recall.

$$F1 = \frac{2 \times P \times R}{P+R} \quad (7)$$

## C. Results and discussion

### 1) Network profiling results

Several tests were carried out to evaluate the success of the proposed IDS framework and determine the accuracy of the machine learning module. During the testing, the threshold parameter of the network profiling service is set to 80% calculated rate difference in the assigned 60 seconds of capture time. Such a large difference from normal transmission rate, in any capture was determined to be a good baseline for our use cases. However, it is important to note that this threshold is able to be configured by the end-user to match their network use cases if their network activity throughput is markedly more volatile, or stable than the SOHO networks we tested the configuration on. Table I present the overall results for the network profiling component. As illustrated in Table I, all the attack scenarios conducted using the network traffic capture replays were detected as out of profile for the devices that have been affected and were correctly identified as such, this yielded a 100% detection success rate for the attacks tested when monitoring out-of- bound network traffic behaviour based on the three network profile assignments against their average profile difference ($\Delta \%$).

TABLE I: Results for the network behaviour profile tests

| Tested scenarios of attack | Out of profile | Detected from profile | Δ(%) |
|---|---|---|---|
| Zero-day exploit | Yes | D | 82.37% |
| DDoS attack with Mirai Botnet | Yes | H, D, W | 98.53 % |
| DDoS attack with Black Energy | Yes | H, D, W | 128.42 % |

| Zeus malware, Linux and windows | Yes | W | 96.70 % |
| Java-RMI backdoor | Yes | D, W | 96.88 % |
| distcc exec backdoor | Yes | D, W | 98.64 % |
| UnrealIRCD backdoor | Yes | D, W | 97.69 % |
| Web Tomcat exploit | Yes | W | 395.52 % |
| Ruby DRb code execution | Yes | D, W | 682.16 % |
| Hydra FTP bruteforce | Yes | D, W | 95.15 % |
| Hydra SSH bruteforce | Yes | D, W | 99.14 % |
| SMTP User Enumeration | Yes | D, W | 93.50 % |
| NetBIOS-SSN | Yes | D, W | 307.39 % |

Network-level attack methods and results are likely to result in false positives (or false alarms) at the expense of accuracy. To test how susceptible our solution is to false positives due to normal user interactions, we conducted testing with benign network traffic scenarios to see what operations, if any would trigger an out of profile response from the profiling system. These tests included extended SSH sessions, downloading and uploading data over FTP, downloading, and uploading data over HTTP; with only one scenario: Copying an extremely large PCAP of 7GB over the network using SCP. This was detected as abnormal by the hourly, daily and weekly profiles with a percentage difference of 281.58%. Out of the 12 scenarios conducted, only one triggered a false positive, meaning the achieved FPR was 8.3%.

*2) IDS System results*

In this work, the machine learning module is implemented by using the MobileNet Convolutional neural network (MobileNetV3). The results for the machine learning component were achieved through the replaying of captured network attack scenarios that were performed on the Cyber-Trust platform. With the PCAP replays to the machine learning component, we were able to assess the metrics of malicious network traffic with quantifiable data; the results of this testing resulted in the following overall statistics. Table II presents the overall results of the conducted attack scenarios, which reached an overall detection accuracy of 98.35%, which is a high rate and met the required accuracy rate in practical use. The accuracy rate has been greatly improved compared to the rates achieved in the design phase (from 95% in [1] to 98%). This is mainly due to the contentious training of the ML module. When observing the results for normal traffic, the average traffic accuracy was in the ranges of 94%-98%, this results in false Positive and False Negative rates of between 1%-4% over full result testing. Comparatively, the results for malicious traffic that was tested results in average traffic accuracy of between 95%-98%, informing a false positive and a false negative rate of between 1%-4% over result testing. These averages do not reflect the best results that were achieved over 100 individual runs of each attack scenario processed by the machine learning component.

While running these proposed testing scenarios of network attack traffic over 100 runs, the best accuracy (A) result achieved was 98.35%, false Positive rate was 0.98% and false Negative rate 0.71%. The precision (P) result was also very high with a rate of 99.3%, which shows strong overall confidence in the pattern recognition process. It is worth noting that in these tests, precision is very important because getting False Negatives (FN), when malware traffic is considered as normal, cost more than False Positives (FP), when normal traffic is considered as malicious traffic. The recall percentage (R) had a result of 99.01%. The F1 value (F1) achieved was 99.16%.

In conclusion, the average performance of both components was considerable. Combining the results of both the network profiling service and the machine learning component, the accuracy rate of both solutions is consistent across our range of testing. The combination of both systems results in an average accuracy rate per iteration of 99.17%. These results were acquired from testing against device exploitation from known common vulnerabilities and high impact botnets that have seen extensive infection in the real world, this speaks to the high efficacy of the solution.

TABLE II: Testing results of the ML module

| Metrics | Obtained values for Normal and malicious traffic | Best results |
|---|---|---|
| Accuracy (%) | ≈ 94% − 98% | **98.35%** |
| FPR (%) | ≈ 1% − 4% | **0.98%** |
| Precision (%) | ≈ 95% − 98% | **99.31%** |
| Recall (%) | ≈ 94% − 98% | **99.01%** |
| F1-score (%) | ≈ 94% − 98% | **99.16%** |

## V. Conclusion and Future Work

In this paper, we have introduced a novel IDS frameworkto identify malicious network traffic in IoT networks by using network profiling and machine learning. This work is done in the context of the Cyber-Trust project [7] andtested on the Cyber-Trust testbed, which hosts a significant number of simulated IoT based home networks (SOHOs). In conclusion, the overall performance of the proposed solution was considerable, this was especially true when considering the results of the machine learning component, which recordeda peak accuracy of 98.35% over 100 tests with only a 0.98% FPR and a 99.31% precision rating. Whereas the overall accuracy rate of the proposed solution is 99.175%. Theseresults were acquired from testing against device exploitation from known common vulnerabilities and high impact botnets that have seen extensive infection in the real world, this speaksto the high efficacy of the solution. The accuracy rate ofthe machine learning component still stands to improve its accuracy rate with further training. It is worth noting that whenit comes to describing future work, tests could be performed toassess whether this model can increase its accuracy with more extensive or alternative forms of binary visualisation training and techniques.

In future work, we intend to improve the results achievedby the ML module, by using more samples for training and testing the classification module, multi-label classification for the malicious traffic behaviour (i.e., keylogger, spyware, etc.) and exploring the possibility of running the ML on low power devices such as the gateway. The profiling service can alsobe improved in future work, allowing for adaptable variance per profile for each device by raising or lowering the out of profile threshold for a device dynamically over time.

## ACKNOWLEDGMENT

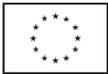
This project has received funding from the European Union's Horizon 2020 research and in-novation programme under grant agreement no. 786698. The work reflects only the authors' view, and the Agency is not responsible for any use that may be made ofthe information it contains.